\documentclass[a4paper,11pt]{article}
\usepackage{pos}
\usepackage{xspace} 

\newcommand{\nch}{\ensuremath{N_{\rm ch}}\xspace}
\newcommand{\nmpi}{\ensuremath{{N}_{\rm mpi}}\xspace}
\newcommand{\pT}{\ensuremath{p_{\rm T}}\xspace}

\title{Multiparton Interactions in pp collisions from Machine Learning}
\ShortTitle{Multiparton Interactions in pp collisions from Machine Learning}

\author*{Erik Zepeda}
\author{Antonio Ortiz}

\affiliation{Instituto de Ciencias Nucleares, Universidad Nacional Autónoma de México,\\
  Apartado Postal 70-543, México Distrito Federal 04510, México}

\emailAdd{eazg@ciencias.unam.mx}
\emailAdd{antonio.ortiz@nucleares.unam.mx}

\abstract{Over the last years, Machine Learning (ML) tools have been successfully applied to a wealth of problems in high-energy physics. In this work, we discuss the extraction of the average number of Multiparton Interactions ($\langle \nmpi \rangle$) from minimum-bias pp data at LHC energies using ML methods. Using the available ALICE data on transverse momentum spectra as a function of multiplicity, we report that for minimum-bias pp collisions at $\sqrt{s} =$ 7\,TeV the average \nmpi is 3.98 $\pm$ 1.01,  which complements our previous results for pp collisions at $\sqrt{s} =$ 5.02 and 13\,TeV. The comparisons indicate a modest energy dependence of $\langle \nmpi \rangle$. We also report the multiplicity dependence of $\nmpi$ for the three center-of-mass energies. These results are qualitatively consistent with the existing ALICE measurements sensitives to MPI, therefore they provide additional experimental evidence of the presence of MPI in pp collisions.}

\FullConference{%
 
 The Ninth Annual Conference on Large Hadron Collider Physics - LHCP2021\\
 7-12 June 2021\\
 Online
}

\begin{document}
\maketitle

\section{Introduction}

The study of Multiparton Interactions (MPI) in pp collisions has recently attracted the attention of the heavy-ion community, because surprisingly, the high-multiplicity pp data unveiled heavy-ion-like features, i.e.  azimuthal anisotropies~\cite{Khachatryan:2010gv}, the enhancement of (multi-)strange hadrons~\cite{ALICE:2017jyt}, as well as radial flow patterns in the transverse momentum (\pT) spectra of identified hadrons~\cite{Acharya:2018orn}.  Besides the hydrodynamical approach~\cite{Bozek:2011if,Nagle:2018nvi}, MPI, offers an alternative possibility to explain the observed phenomena. For instance, color reconnection and MPI can mimic radial flow patterns in pp collisions~\cite{Ortiz:2013yxa}. In this direction, we have proposed the extraction of the MPI activity from minimum-bias pp data using Machine Learning (ML) methods~\cite{Ortiz:2020rwg,Ortiz:2021peu}. In this contribution, we summarize the main results including the multiplicity dependence of the average number of MPI extracted from the available ALICE data at the LHC~\cite{Acharya:2019mzb,Acharya:2018orn}.

\section{Analysis}

Our approach relies on a multivariate regression technique based on Boosted Decision Trees (BDT). The study is conducted using the Toolkit for Multivariate Analysis (TMVA) framework which provides a ROOT-integrated ML environment for the processing and parallel evaluation of multivariate classification and regression techniques~\cite{Voss:2007jxm}. The training is performed using pp collisions at $\sqrt{s}=13$\,TeV simulated with PYTHIA~8.244~\cite{Sjostrand:2014zea} tune 4C~\cite{Corke:2010yf}. The choice of the input variables is based on their correlation with \nmpi~\cite{Cuautle:2015fbx}. We consider the event-by-event average transverse momentum and the mid-pseudorapidity charged particle multiplicity ($N_{\rm ch}$).

\begin{figure*}[h!]
\begin{center}
\includegraphics[width=11.0cm]{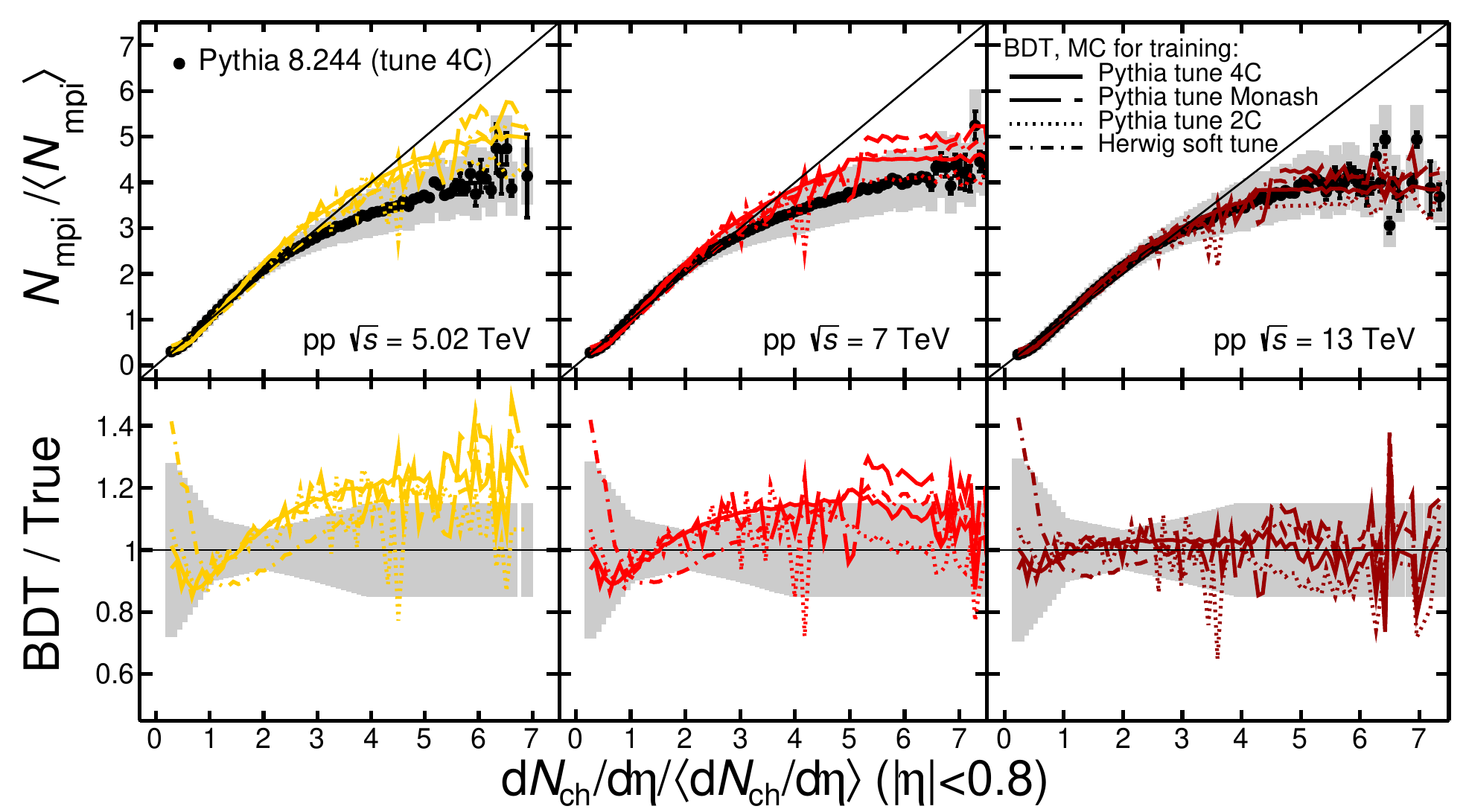}
\caption{Monte Carlo closure test using pp collisions at $\sqrt{s}=5.02$ (left), 7 (middle) and 13\,TeV (right) simulated with PYTHIA~8 tune 4C. The top panels display the self normalized average number of MPI as a function of the self-normalized mid-pseudorapidity charged particle multiplicity. Ratios between ML results and the true values given by PYTHIA are shown in the bottom panels.}
\label{fig:1}
\end{center}
\end{figure*}

\vspace{-20pt}

The systematic uncertainties take into account a variation of the PYTHIA~8 tune, as well as the MPI and hadronization model. To this end, tunes: 2C, 4C and Monash~2013 were used for training, and the effects of MPI and hadronization was investigated using the Monte Carlo (MC) generator HERWIG~7.2~\cite{Bellm:2019zci} for training.
Figure~\ref{fig:1} shows the correlation between the self normalized number of MPI ($\nmpi / \langle \nmpi \rangle$) and the self normalized mid-pseudorapidity charged particle multiplicity ($\nch / \langle \nch \rangle$) in pp collisions at $\sqrt{s}=5.02$, 7 and 13\,TeV. For $\nch / \langle \nch \rangle<3$, the self normalized \nmpi increases linearly with the event multiplicity. On the other hand, for higher multiplicities, we observe a deviation of the self normalized \nmpi with respect to the linear trend. The Figure~\ref{fig:1} also displays the results obtained from regression (lines), and shows that using ML-based regression, one can recover the energy and multiplicity dependence.

\begin{figure*}[h!]
\begin{center}
\includegraphics[width=9.0cm]{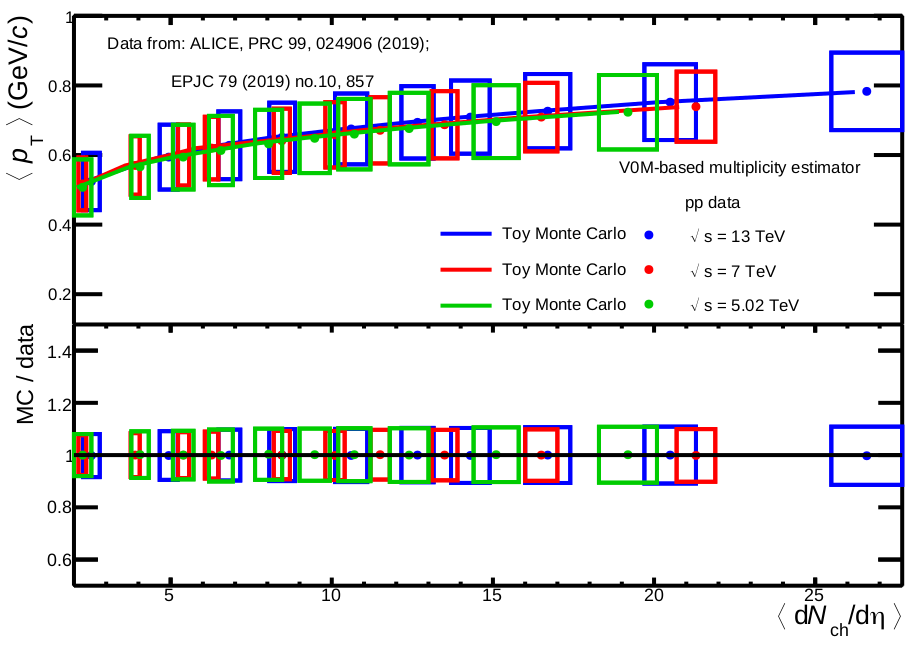}
\caption{Mean transverse momentum as a function of the average charged-particle multiplicity density in pp collisions at $\sqrt{s}=5.02$, 7 and 13\,TeV. In the top panel ALICE data~\cite{Acharya:2019mzb,Acharya:2018orn} (solid markers) are compared with results from the Toy MC (solid lines). Bottom panel displays ratios between Toy MC and the data.}
\label{meanptvsN}  
\end{center}
\end{figure*}

\vspace{-20pt}

Regarding the analysis using available data, we built a toy MC in order to get the correlation between the event-by-event $\langle p_{\rm T} \rangle$ and \nch. Figure~\ref{meanptvsN} displays the mean transverse momentum as a function of the average charged-particle multiplicity density in pp collisions at $\sqrt{s}=5.02$, 7 and 13\,TeV. Within uncertainties, the toy MC reproduces the correlation between the $\langle p_{\rm T} \rangle$ and $\langle d \nch/d\eta \rangle$. In our approach, the event-by-event information produced by the toy MC was processed with the trained BDT in order to extract the MPI activity associated with the data.

\section{Results}

\begin{figure*}[h!]
\begin{center}
\includegraphics[width=7.5cm]{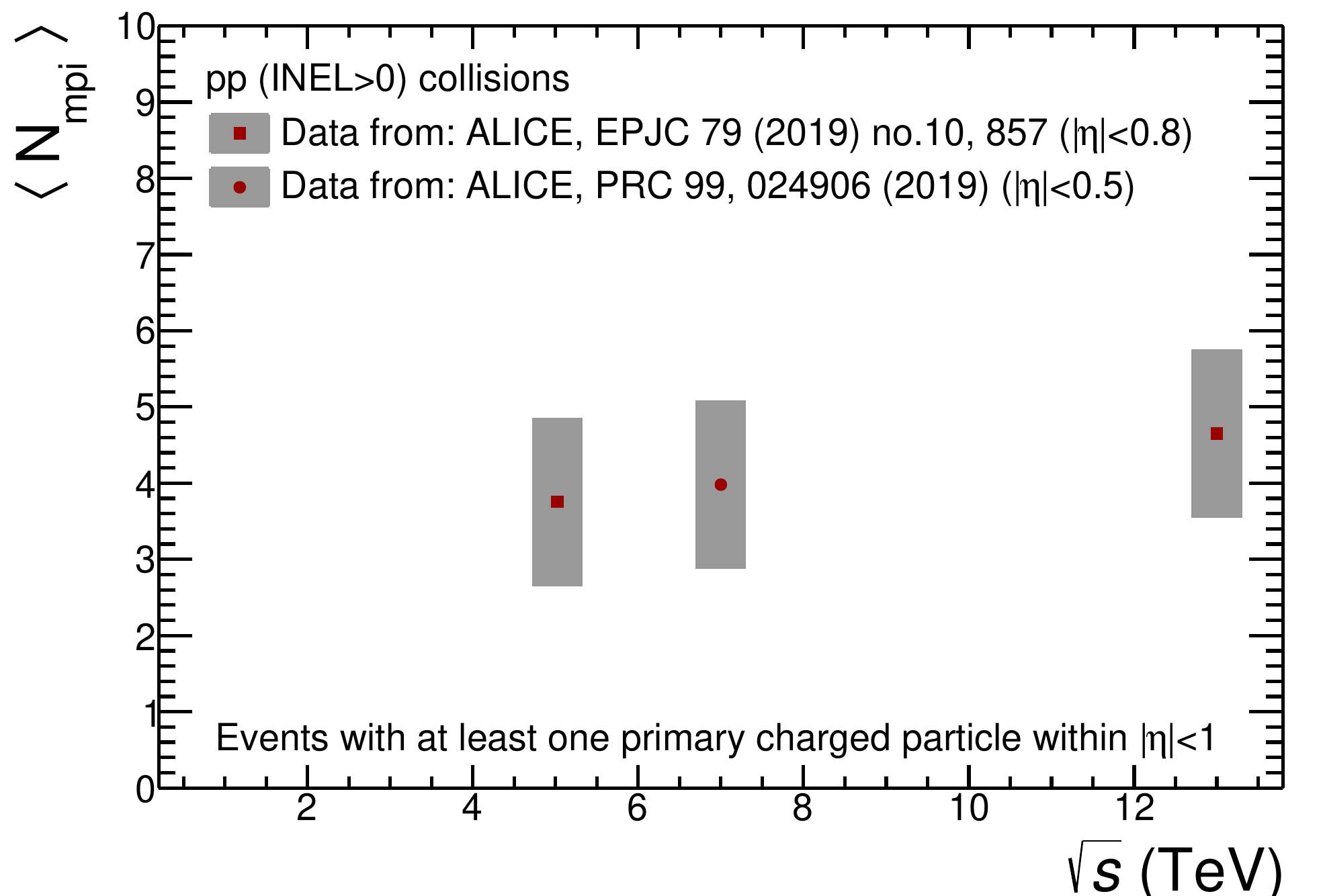}
\caption{Average number of MPI as a function of the center-of-
mass energy. Results for pp collisions at $\sqrt{s}=7$\,TeV, are compared to those for pp collisions at $\sqrt{s}=5.02$ and 13\,TeV reported in~\cite{Ortiz:2020rwg}.}
\label{fig:2}
\end{center}
\end{figure*}

\begin{figure*}
\begin{center}
\includegraphics[width=7.5cm]{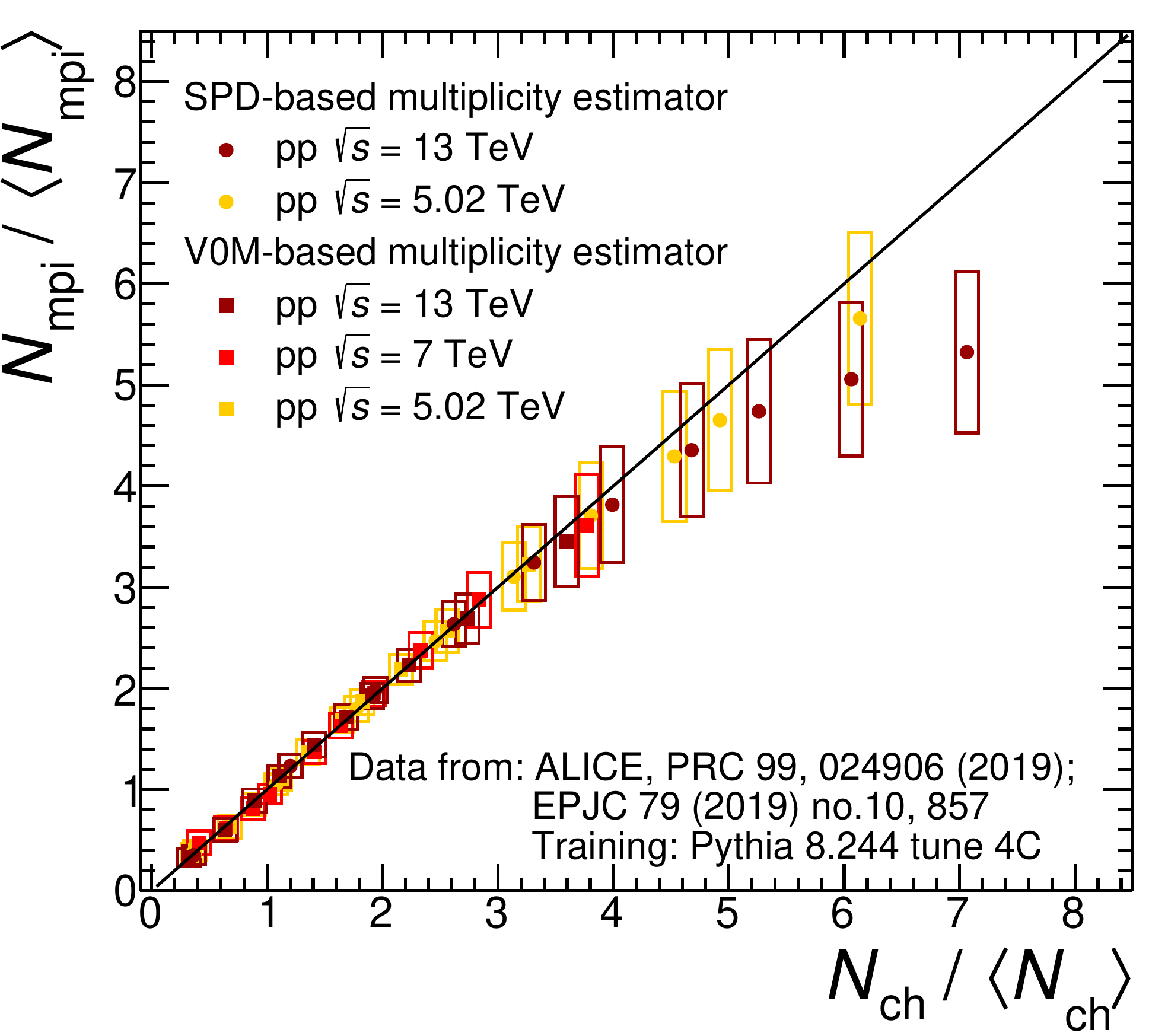}
\caption{The self normalized average number of Multiparton Interactions as a function of the self normalized mid-pseudorapidity charged particle multiplicity is shown for pp collisions at $\sqrt{s}=5.02$, 7 and 13\,TeV.}
\label{fig:3}  
\end{center}
\end{figure*}

Using the ALICE data from pp collisions at $\sqrt{s}=7$\,TeV~\cite{Acharya:2018orn}, we extract the average number of MPI, which is found to be $\langle N_{\rm mpi} \rangle = 3.98 \pm 1.01$. Figure~\ref{fig:2} displays the average number of MPI as a function of the center-of-mass energy, for pp collision at $\sqrt{s}$= 5.02, 7 and 13\,TeV. We obtain a regression value which is above unity, therefore, our results support the presence of MPI in pp collisions. We also observe a modest energy dependence, which is similar to that predicted by PYTHIA~8~\cite{Ortiz:2020rwg}. A similar finding has been discussed in Ref.~\cite{Ortiz:2021gcr}, where the energy dependence of underlying-event observables has been reported. Figure~\ref{fig:3} displays the self normalized number of MPI  ($\nmpi / \langle \nmpi \rangle$) as a function of the self-normalized mid-pseudorapidity charged-particle multiplicity ($\nch / \langle \nch \rangle$) in pp collisions at $\sqrt{s}=5.02$, 7 and 13\,TeV from ALICE data. We observe that for $\nch  <3 \langle \nch \rangle $ the self normalized \nmpi increases linearly with the event multiplicity. Regarding higher multiplicities, we observe a deviation of the self normalized \nmpi with respect to the linear trend. This result qualitatively agrees with PYTHIA~8 (see figure~\ref{fig:1}).

\vspace{10cm}

\vspace{-20pt}

\section{Conclusions}

We report the extraction of the average number of MPI from pp data at the LHC energies. We have found $\langle N_{\rm mpi} \rangle = 3.98 \pm 1.01$ for pp collisions at $\sqrt{s}=$ 7 TeV. The comparisons with our previous results for pp collisions at $\sqrt{s}=5.02$ and 13\,TeV indicate a modest energy dependence of \nmpi. This result provide experimental evidence of the presence of MPI in hadronic interactions. In addition, we also report the multiplicity dependence of \nmpi for the three center-of-mass energies. Our results are fully consistent with the so-called ``mini-jet analysis'' of ALICE~\cite{Abelev:2013sqa}, and suggest that high multiplicities (at mid-pseudorapidity) can only be reached by selecting events with many high-multiplicity jets.

\section*{Acknowledgments}

Authors acknowledge Antonio Paz for providing the simulations with HERWIG~7.2. Support for this work has been received from CONACyT under the Grant No. A1-S-22917. E. Z. acknowledges the fellowship of CONACyT.

\bibliographystyle{unsrt}
\bibliography{biblio}

\begin{thebibliography}{10}

\bibitem{Khachatryan:2010gv}
Vardan Khachatryan et~al.
\newblock {Observation of Long-Range Near-Side Angular Correlations in
  Proton-Proton Collisions at the LHC}.
\newblock {\em JHEP}, 09:091, 2010.

\bibitem{ALICE:2017jyt}
Jaroslav Adam et~al.
\newblock {Enhanced production of multi-strange hadrons in high-multiplicity
  proton-proton collisions}.
\newblock {\em Nature Phys.}, 13:535--539, 2017.

\bibitem{Acharya:2018orn}
Shreyasi Acharya et~al.
\newblock {Multiplicity dependence of light-flavor hadron production in pp
  collisions at $\sqrt{s}$ = 7 TeV}.
\newblock {\em Phys. Rev. C}, 99(2):024906, 2019.

\bibitem{Bozek:2011if}
Piotr Bozek.
\newblock {Collective flow in p-Pb and d-Pd collisions at TeV energies}.
\newblock {\em Phys. Rev.}, C85:014911, 2012.

\bibitem{Nagle:2018nvi}
James~L. Nagle and William~A. Zajc.
\newblock {Small System Collectivity in Relativistic Hadronic and Nuclear
  Collisions}.
\newblock {\em Ann. Rev. Nucl. Part. Sci.}, 68:211--235, 2018.

\bibitem{Ortiz:2013yxa}
Antonio Ortiz, Peter Christiansen, Eleazar Cuautle~Flores, Ivonne
  Maldonado~Cervantes, and Guy Pai\'c.
\newblock {Color Reconnection and Flowlike Patterns in $pp$ Collisions}.
\newblock {\em Phys. Rev. Lett.}, 111(4):042001, 2013.

\bibitem{Ortiz:2020rwg}
Antonio Ortiz, Antonio Paz, Jos\'e~D. Romo, Sushanta Tripathy, Erik~A. Zepeda,
  and Irais Bautista.
\newblock {Multiparton interactions in $pp$ collisions from machine
  learning-based regression}.
\newblock {\em Phys. Rev. D}, 102(7):076014, 2020.

\bibitem{Ortiz:2021peu}
Antonio Ortiz and Erik~A. Zepeda.
\newblock {Extraction of the multiplicity dependence of multiparton
  interactions from LHC pp data using machine learning techniques}.
\newblock {\em J. Phys. G}, 48(8):085014, 2021.

\bibitem{Acharya:2019mzb}
Shreyasi Acharya et~al.
\newblock {Charged-particle production as a function of multiplicity and
  transverse spherocity in pp collisions at $\sqrt{s} =5.02$ and 13 TeV}.
\newblock {\em Eur. Phys. J. C}, 79(10):857, 2019.

\bibitem{Voss:2007jxm}
H.~Voss, Andreas Hocker, J.~Stelzer, and F.~Tegenfeldt.
\newblock {TMVA, the Toolkit for Multivariate Data Analysis with ROOT}.
\newblock {\em PoS}, ACAT:040, 2007.

\bibitem{Sjostrand:2014zea}
Torbj\"orn Sj\"ostrand, Stefan Ask, Jesper~R. Christiansen, Richard Corke,
  Nishita Desai, Philip Ilten, Stephen Mrenna, Stefan Prestel, Christine~O.
  Rasmussen, and Peter~Z. Skands.
\newblock {An introduction to PYTHIA 8.2}.
\newblock {\em Comput. Phys. Commun.}, 191:159--177, 2015.

\bibitem{Corke:2010yf}
Richard Corke and Torbjorn Sjostrand.
\newblock {Interleaved Parton Showers and Tuning Prospects}.
\newblock {\em JHEP}, 03:032, 2011.

\bibitem{Cuautle:2015fbx}
Eleazar Cuautle, Antonio Ortiz, and Guy Paic.
\newblock {Effects produced by multi-parton interactions and color reconnection
  in small systems}.
\newblock {\em Nucl. Phys. A}, 956:749--752, 2016.

\bibitem{Bellm:2019zci}
Johannes Bellm et~al.
\newblock {Herwig 7.2 release note}.
\newblock {\em Eur. Phys. J. C}, 80(5):452, 2020.

\bibitem{Ortiz:2021gcr}
Antonio Ortiz.
\newblock {Energy dependence of underlying-event observables from RHIC to LHC
  energies}.
\newblock 8 2021. arXiv:2108.08360 [hep-ph].

\bibitem{Abelev:2013sqa}
Betty Abelev et~al.
\newblock {Multiplicity dependence of two-particle azimuthal correlations in pp
  collisions at the LHC}.
\newblock {\em JHEP}, 09:049, 2013.

\end{thebibliography}

\end{document}